# Analytical Energy Formalism and Kinetic Effects of Grain Boundary: A Case Study of Graphene

Cheng-yan Liu,[†,‡] Zhi-ming Li,[†] Xin-gao Gong[*,†]

[†]Key Laboratory for Computational Physical Sciences (MOE), State Key Laboratory of Surface Physics, Department of Physics, Fudan University, Shanghai 200433, China

[‡]Key Laboratory for Special Functional Materials of Ministry of Education Collaborative Innovation Center of Nano Functional Materials and Applications, School of Materials Science and Engineering, Henan University, Kaifeng, Henan 475001, China

E-mail:xggong@fudan.edu.cn

Grain boundaries (GBs), an important constituent of polycrystalline materials, have a wide range of manifestion and significantly affect the properties of materials. Fully understanding the effects of GBs is stalemated due to lack of complete knowledge of their structures and energetics. Here, for the first time, by taking graphene as an example, we propose an analytical energy functional of GBs in angle space. We find that an arbitrary GB can be characterized by a geometric combination of symmetric GBs that follow the principle of uniform distribution of their dislocation cores in straight lines. Furthermore, we determine the elusive kinetic effects on GBs from the difference between experimental statistics and energy-dependent thermodynamic effects. This study not only presents an analytical energy functional of GBs which could also be extended to other two-dimensional materials, but also sheds light on understanding the kinetic effects of GBs in material synthesizing processes.

KEYWORDS: Grain boundaries, thermodynamic effects, kinetic effects

A grain boundary (GB) is a stitching interface between two grains with different orientations. GBs are an important aspect of polycrystalline materials. These materials have been the subject of intensive research for many decades. The existence of GBs breaks the translation symmetry of crystals and thus may affect the physical properties in either a beneficial or detrimental way. The formation of GBs results from an unmanageable growth process with both kinetic and thermodynamic contributions.

Currently, understanding and utilizing the properties of GBs are stalemated by their complicated nature. Although great efforts have been made both experimentally[1-6] and theoretically[6-13], the relation between atomic configuration and grain-boundary energy has not been obtained. Computational science seems a feasible way to screen out the energetically favorable GB structures through an ergodic search, but it is laborious and unrealistic for each system of interest. If the constructing mechanism of energetically favorable GBs is understood, the corresponding relation between energy and structure is likely to be obtained. The analytical Read-Shockley formula expressed in macroscopic parameters of shear modulus $G$ and Poisson's ratio $v$ [14] attempts to present this relation, but it only works at very small grain-boundary angles[8,15]. A previous theoretical study has achieved great success in the qualitative determination of GB paths and judgement of grain-boundary energy in the graphene system[10]. Up to now, a quantitative description of the relation between grain-boundary energy and misorientation angle is still hindered by large angle cases, since the dense dislocations of large angle GBs are not suitable for the energy integral of slide planes that are adapted to the formula described by the macroscopic parameters[14]. Finally, there is a point of view that no general and useful criterion for grain-boundary energy can be established in a simple geometric framework if there is no consideration of the atomic structure and the details of bonding at the interface[16].

In this letter, for the first time we present a general analytical functional of grain-boundary energy without distinguishing large GB angles from small GB angles for two-dimensional materials. The establishment of this analytical functional is based on two physical understandings: (1) any asymmetric GB can be decomposed into the vector sum of two symmetric GBs, and (2) the symmetric GB are constructed such that basic dislocation cores are aligned as uniformly as possible to minimize inter-grain stress. Among two-dimensional materials, graphene is the best candidate to test our functional owing to the advantages of its simple atomic structure and widely studied GB samples[8,15-17], which are helpful to characterize the dominant features of GBs. On the premise of understanding GB energy in entire angle space, the elusive kinetic effects for the first time are determined for the graphene system from the difference between our theoretical predictions and experimental statistics.

The atomic configuration of a GB is determined by the competition between two grains on both sides of boundary, which drive the interface atoms to conform to their own lattice orientations as much as possible. A symmetric GB, as the name implies, possesses mirror symmetry and exhibits the shortest straight dislocation route due to the equivalent stress on both sides of boundary based on the GB theory of solids[18]. However, for an asymmetric GB, the asymmetric grains on either side lead to an irregular, sinuous route to minimize the inter-grain stress.

A GB in a two-dimensional system can be uniquely described by two independent angle variables $\theta_M$ and $\theta_L/2$, as schematically shown in Fig. 1a. The variable $\theta_M$ represents the grain-boundary angle formed by two grain orientations marked by gray and green dashed lines. The variable $\theta_L/2$ represents the asymmetric angle formed by the bisector of $\theta_M$ (red solid line) and GB direction (black solid line). The special condition of $\theta_M = 0$ or $\theta_L/2 = 0$ leads to a perfect lattice or a symmetric GB. Asymmetric GBs ($\theta_M \neq 0$, $\theta_L/2 \neq 0$) are quite different from symmetric GBs ($\theta_M \neq 0$, $\theta_L/2 = 0$) which tend to be straight due to the grain symmetry. Asymmetric GBs energetically optimize their internal stress by decomposing into the vector sum of two symmetric sub-GBs.

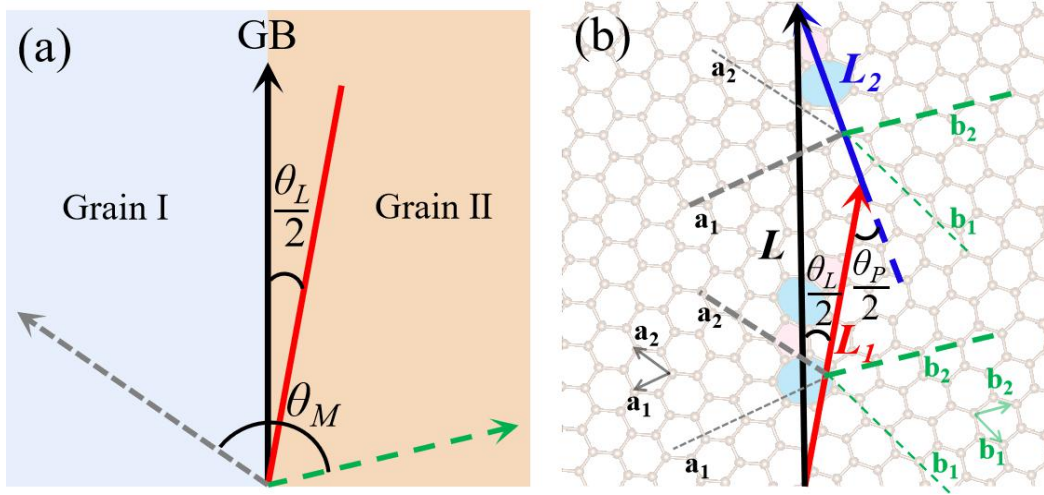

FIG 1. Schematic model of an asymmetric GB. (a) Grain I and Grain II with different grain orientations, marked by gray and green dashed lines. Two grains stitching together form a GB marked by black solid line. The red solid line bisects the grain-boundary angle $\theta_M$ with an asymmetric angle of $\theta_L/2$ respecting to the black solid line of GB orientation. (b) The vector of asymmetric GB with periodic unit length L is decomposed into the vectors of two symmetric sub-GBs represented by $L_1$ and $L_2$ respectively. Periodic length L is determined by the commensurability width of two grains along the GB direction. One is set up along the bisector direction marked by $L_1$ and the other is marked by $L_2$. $L_2$ rotates a half of periodic rotation symmetry angle $\theta_P/2$ respecting to $L_1$, which is also the bisector direction on another end of GB (for graphene, $\theta_P = 60°$). Arrows $a_1$ and $a_2$ are the lattice vectors of grain I. Arrows $b_1$ and $b_2$ are the lattice vectors of grain II. $L_1$ and $L_2$ bisect the angles formed by $(a_2, b_2)$ and $-(a_1, b_2)$, respectively.

Geometrically, the atomic structure of a specific ($\theta_M$, $\theta_L$) GB cannot be uniquely determined only by confining the starting and ending points of a dislocation route between both ends of GB's periodic unit. The dislocation route consists of the locations where grain orientation changes from one direction to the other. However, the routes of energetically favorable GBs, which dominantly occur in materials, are almost unique, and they possess the merit of maximally releasing the interface stress with minimum energy consumption. Viewing dislocation cores as the basic units of GBs, dislocation cores are a result of the matched competition between two grains.

Therefore, any GB can be regarded as the splicing of the local symmetrical GBs with dislocation cores as the smallest symmetric centers. In order to minimize the GB energy, dislocation cores must be arranged according to the shortest path. Therefore, we reasonably speculate that a symmetric GB has a straight dislocation route, and an energetically favorable asymmetric GB is the vector sum of different symmetric GBs. Additionally, a symmetric GB also complies with the constructing ideology of an asymmetric GB, which can be regarded as the combination of two identical symmetric sub-GBs.

Here, we present the decomposition model schematically based on an asymmetric graphene GB as shown in Fig. 1b. The symmetric sub-GBs of red line $L_1$ and blue line $L_2$ are both the bisectors of grain-boundary angles starting from the two ends of periodic length $L$ of the asymmetric GB. Analysis of rotational symmetry indicates that asymmetric angles on both ends of an asymmetric GB are the values of $\theta_L/2$ and $(\theta_P - \theta_L)/2$, where $\theta_P$ is the periodic angle of rotational symmetry in the lattice, such as $\theta_P = 60°$ for hexagonal graphene. Therefore, an asymmetric GB ($L$) can be successfully characterized by two symmetric sub-GBs ($L_1$ and $L_2$) which form along the bisector of grain-boundary angle $\theta_M$ and $\theta_P/2$ anticlockwise rotation relative to it, respectively. Therefore, according to the geometric relation of triangle side length ($L_1$, $L_2$ and $L$ as shown in Fig. 1b), grain-boundary energy as a function of variables $\theta_M$ and $\theta_L$ can be derived as shown in eq. 1. The detailed derivation procedure is presented in part 1 of supplement material (SM)[19].

$$E_f(\theta_M, \theta_L) = \frac{1}{\sin\left(\frac{\theta_P}{2}\right)} \left[ \sin\left(\frac{\theta_P - \theta_L}{2}\right) f_b(\theta_M) + \sin\left(\frac{\theta_L}{2}\right) f_b(\theta_P - \theta_M) \right] \quad (1)$$

Where $E_f$ is the grain-boundary energy as a function of the variables $\theta_M$ and $\theta_L$, and $f_b$ is the grain-boundary energy per unit length of the symmetric GB. The first term and second term in the square brackets of eq. 1 represent the energy of the decomposed symmetric sub-GBs labeled $L_1$ and $L_2$, respectively. If the asymmetric angle $\theta_L/2$ is equal to 0, the eq. 1 reduces to the energy of the symmetric GB. Since $\theta_P$ is a constant, 60° for a hexagonal lattice, the grain-boundary energy just depends on the energy of the symmetric GB represented by functional $f_b$. Usually, the number of symmetric GBs is discrete and limited. One can numerically calculate all the symmetric GBs without much difficulty, alternatively, as we did below, one can get the energy functional $f_b$ by fitting the calculated symmetric GB energies. In this way, the energy of an arbitrary GB can be easily obtained from eq. 1.

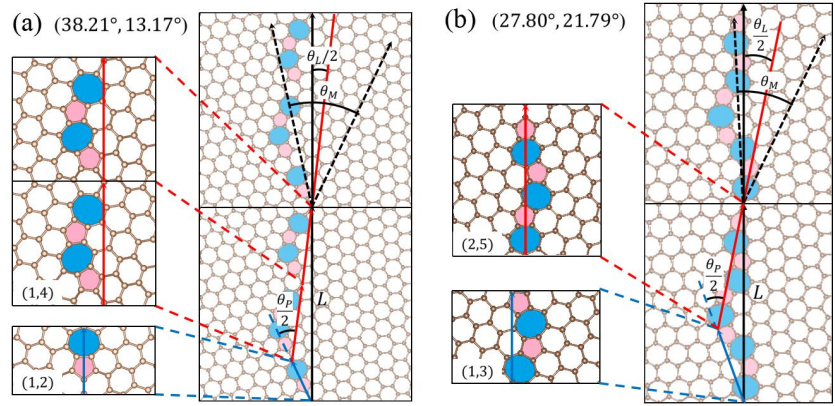

FIG. 2. The decomposition model of asymmetric graphene GBs. (a) Asymmetric (38.21°, 13.17°) GB is decomposed into symmetric (1, 4) and (1, 2) sub-GBs. The description of (1, 4) characterizes the symmetric GB with the vector sum of one lattice vector **a** and four times lattice vector **b**. The same meaning applies to the following descriptions and SM. (b) Asymmetric (27.80°, 21.79°) GB is decomposed into symmetric (2, 5) and (1, 3) sub-GBs. $\theta_P/2$ is 30° because of the 60° rotational periodic angle for hexagonal graphene.

Here, we take six samples of asymmetric GBs to verify our decomposition model by comparing the grain-boundary energies from direct first-principles calculations and from eq. 1. Fig. 2a and 2b show two asymmetric GBs with the variables ($\theta_M$, $\theta_L$) which are (38.21°, 13.17°) and (27.80°, 21.79°), respectively. We can see that one unit of asymmetric (38.21°, 13.17°) GB can be decomposed into double units of symmetric (1, 4) sub-GB and one unit of symmetric (1, 2) sub-GB as shown in Fig. 2a.

Two sub-GB orientations, marked by red and blue lines, comply with the decomposition model described in Fig. 1a and 1b. Similarly, asymmetric (27.80°, 21.79°) GB is decomposed into symmetric (2, 5) and (1, 2) sub-GBs as shown in Fig. 2b. The decomposition models of another four asymmetric GBs, (38.21°, 32.20°), (46.83°, 21.79°), (21.79°, 2.79°), and (38.21°, 38.21°), are shown in Fig. S8. We perform first-principles calculations for these six asymmetric GBs as well as their corresponding symmetric GBs. The details of the calculations are described in SM parts 5 and 6. All the calculated grain-boundary energies, from the direct calculation and from Eq. 1, are listed in Table 1. One can see that the errors between direct calculations and predictions by eq. 1 are below 4%. Simultaneously, we carefully re-optimize the atomic structures of (6, 5)|(9, 1) GBs with (27.80°, 21.79°) reported in ref. 10 as shown in Fig. S9. (6, 5)|(9, 1) represents an asymmetric GB formed by two grains with vector sum of (6, 5) and (9, 1), respectively. We would like to point out that, for the asymmetric (27.80°, 21.79°) GB, our present calculations with carefully optimized atomic structure as shown in Fig. 2b, both from direct calculation and decomposition model, predict 0.06 eV lower than that of the most stable structure reported in ref. 10. These results strongly support the reliability of our GB decomposition model.

In order to get the energy functional $f_b$ of the symmetric GB used in the square bracket of eq. 1, we construct the atomic structure of the symmetric GB based on the principle of uniform arrangement of dislocation cores on a straight line. For graphene, five kinds of dislocation cores of different symmetric grain-boundary angles are refined from our previous study[20] and elsewhere in the literature[8, 10, 11, 21-27] (details in SM part 3). Here, we construct 62 symmetric GBs with 37 unique grain-boundary angles distributed 0 ~ 60° as shown in Fig S3-7, and all the formation energies are shown in Fig. S2d. By fitting the lowest energies of symmetric GBs, the energy functional $f_b$ for a symmetric GB is obtained as shown in eq. S4. The energy curve with the fitting error smaller than 0.04 eV/nm as a function of grain-boundary angle is shown in Fig. 3a.

Table 1. The grain-boundary energies (eV/nm) of six GB structures from direct calculations of DFT and from predictions of eq. 1. The energies of symmetric GBs used in eq. 1 come from direct DFT calculations (DFT sym GB) and from fitting a quadratic function (Fitting sym GB), respectively. Atomic structures of the first two samples are shown in Fig. 2a and 2b. The other four structures of GBs are presented in S8 a-d[19]. Below 4% error between direct the DFT calculations and the derivations of eq. 1 indicates that our decomposition model is reliable.

| ($\theta_M$, $\theta_L$) | DFT asym GB | Eq. 1 predictions | | Error % | |
| --- | --- | --- | --- | --- | --- |
| | | DFT sym GB | Fitting sym GB | DFT sym GB | Fitting sym GB |
| (38.21°, 13.17°) | 4.60 | 4.57 | 4.56 | -0.65 | -0.87 |
| (27.80°, 21.79°) | 4.33 | 4.35 | 4.34 | 0.46 | 0.23 |
| (38.21°, 32.20°) | 4.58 | 4.52 | 4.52 | -1.31 | -1.31 |
| (46.83°, 21.79°) | 3.91 | 3.97 | 4.06 | 1.53 | 3.83 |
| (21.79°, 21.79°) | 4.54 | 4.48 | 4.48 | -1.32 | -1.32 |
| (38.21°, 38.21°) | 4.54 | 4.48 | 4.48 | -1.32 | -1.32 |

As soon as we get the energy functional $f_b(\theta_M)$ of symmetric GBs as shown in Fig. 3a, we determine the energy landscape of a GB within the space of symmetric grain-boundary angle $\theta_M$ and asymmetric angle $\theta_L$ as shown in the side view of Fig. 3b and top view of Fig. 3c. We can see that the variation of grain-boundary energy along the axis of $\theta_M$ is more prominent than that along the axis of $\theta_L$, since the grain-boundary energy significantly depends on the magnitude of its Burger vector expressed by $2\sin(\theta_M/2)$ and is independent of $\theta_L$. The energy landscape as shown in Fig. 3b is very well in agreement with what is shown in Fig. 4a of Ref. 13 which is directly depicted from 79,000 numerical simulations. This strongly confirms that our analytical energy functional of eq. 1 can accurately characterize the energetics of GBs in graphene.

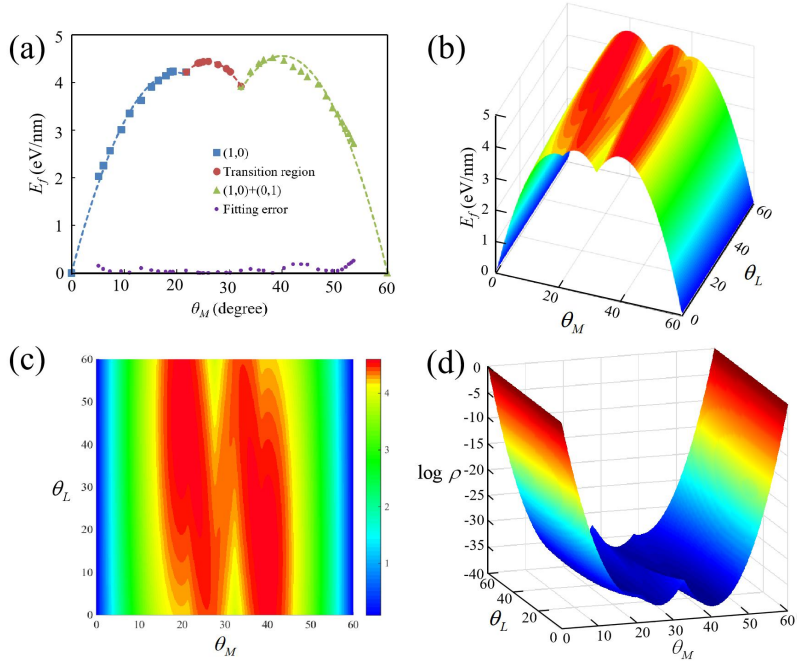

FIG. 3 Grain-boundary energy (eV/nm) and thermodynamic distribution probability of a graphene GB. (a) The grain-boundary energy of symmetric GBs as a function of the grain-boundary angle $\theta_M$. The dispersed values come from the lowest energies of symmetric GBs which are constructed directly based on the constructing principle (details in Fig. S2d and S3-7). The solid line is the fitted quadratic function, and the purple points represent the fitting error, which is smaller than 0.04 eV/nm. (b) Grain-boundary energy landscape as a function of $\theta_M$ and $\theta_L$ based on eq. 1. (c) The top view of (b). (d) The logarithm of probability $\rho$ as a function of $\theta_M$ and $\theta_L$ under thermodynamic effects for graphene GBs.

The thermodynamic effects of GBs come from the contribution of atomic vibration entropy and GB formation energy. The energy contributed by entropy is negligible compared with the formation energy of GBs. We examine the situations of 0°, 21.79° and 16.43° symmetric GBs, which correspond to the 0 GB energy of perfect lattice, almost the maximum value of grain-boundary energy, and the value in the middle of them, respectively, as shown in SM part 9. We can see that the thermodynamic energy contributed by entropy is very small at 1000 K, which does not exceed 12.67% compared to the GB formation energies. The increase of energy induced by entropy is also a negligible quantity in the studies of defect and GB formation energies[28,29]. Therefore, the thermodynamic probabilities, as shown in Fig. 3d, can be calculated

directly through canonical ensemble as

$$\rho = \frac{1}{Z} e^{-\frac{1}{k_B T} E_f(\theta_M, \theta_L)}, \qquad (2)$$

where $\rho$ is the occurrence probability of a graphene GB with the grain-boundary energy of $E_f(\theta_M, \theta_L)$, and $Z$ and $k_B$ are the partition function and Boltzmann constant, respectively. Growth temperature $T$ is selected as 1000 K, referring to the chemical vapor deposition (CVD) condition of graphene. The logarithmic thermodynamic probability in angle space of $\theta_M$ and $\theta_L$ are shown in Fig. 3d.

The formation of GBs is affected not only by thermodynamic effects but also by kinetic effects. Quantitatively determining the kinetic effects is a very difficult task due to elusive impacts in the synthesizing process. Here, for the first time we reveal the kinetic effects on the formation of graphene GBs from the difference between experimental statistics[5,30] and our thermodynamic prediction as shown in Fig. 4. We find that experimental statistics of GB probability is only characterized by grain-boundary angle $\theta_M$ neglecting the asymmetric angle $\theta_L$, which leads to the experimental results presented within a half of periodic angle $\theta_M/2$ (30° for graphene). In addition, large angle GBs are selected deliberately for statistics ($\theta_M$ from 15° to 30°) in experiments owing to the prominently distinguishable feature of their dense dislocation cores. In order to present matching theoretical data, we absorb the variable of asymmetric angle $\theta_L$ by integrating eq. 1 and consider the same range of grain-boundary angle (15° ~ 30°) folded from 15° ~ 45° with 30° as the symmetric center. In this way, the theoretical probabilities of GBs in the range of 15° ~ 30° are obtained as shown in green histograms of Fig. 4a.

In Fig. 4a, we can see that two different experimental statistics shown in red and blue histograms have similar variation trends, and the difference between them reflects the dependence of kinetic effects on synthesizing conditions during the formation of GBs. The overall trend of our theoretical calculations is consistent with the experimental statistics, which presents a relatively high probability in the vicinity of 15° and 28° and a relatively low probability for the rest of the angles. The reason is that the probability of the folded 15° in Fig. 4a comes from the probability superposition of 15° and 45° which have the lowest formation energies near the two ends of the concerned energy curve as shown in Fig. 3a. Viewing their atomic structures, both of them have sparser dislocation cores than other GBs in the range of 15° ~ 45° as shown in Fig. S3 and S6. For the probability of the folded 28° in Fig. 4a, it comes from the probability superposition of 28° and 32° as shown in Fig. 3a. Since the 32° is very close to the transition point 32.2° between near-(1, 0)+(0, 1) and (1, 0)+(0, 1) dislocation cores, it has a lower energy than that of its vicinity. The near-(1, 0)+(0, 1) and (1, 0)+(0, 1) dislocation cores are presented in Fig. S2. Therefore, 15° and 28° are the two extreme points of energy within the folded 15° ~ 30°, which possess high probabilities compared to those of other angles as shown in Fig. 4a.

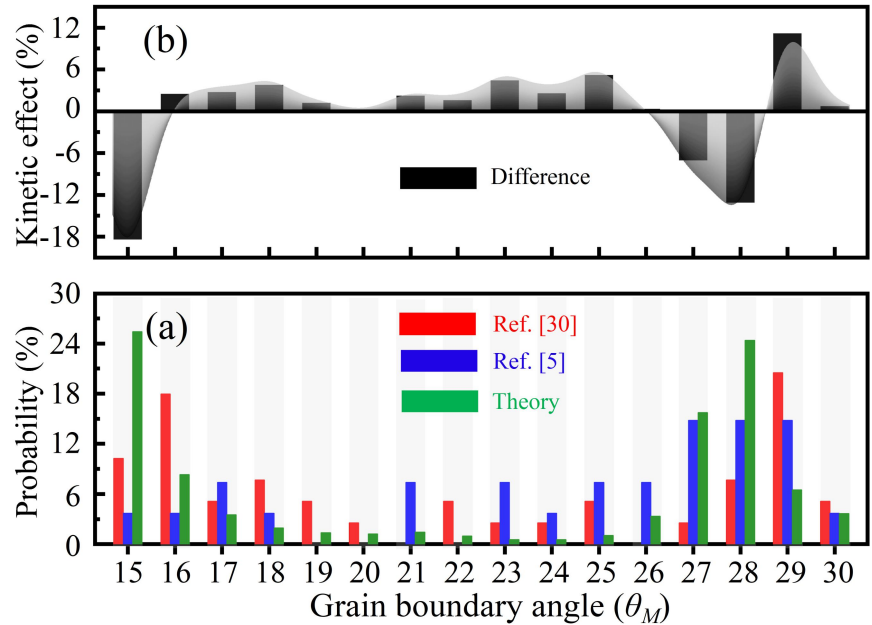

FIG. 4 Kinetic effects on graphene GBs. (a) The experimental statistics of GB probabilities shown in blue[5] and red[30] histograms and the theoretical thermodynamic prediction shown in the green histogram. (b) Kinetic effects revealed by the difference between the probabilities of thermodynamic prediction and the average value of experimental statistics[5,30]. Negative values and positive values indicate the weakening and enhancing effects of kinetics, respectively. The continuous kinetic effects are represented by Gaussian broadening as shown in shadow area.

Since experimental statistics result from both kinetic and thermodynamic effects, we extract kinetic effects from the difference between the thermodynamic probability and the average of experimental results. The kinetic effects obtained is displayed in black histograms and shadow area of Gaussian broadening as shown in Fig. 4b. We can see that kinetic effects make the experimental statistics smoother than thermodynamic calculations, reducing the probability strongly near the 15° and 28° where the thermodynamic maximums are, while increasing the probability for the rest of the GB angles where the thermodynamics predicts it to be low.

To understand this phenomenon, we can imagine the extreme case that if there is no difference among formation energies of GBs or very large kinetic effects which make the formation energy negligible, the probabilities of GBs in all grain-boundary angles must be identical. Since experiments need to obtain a large graphene film without too many dislocations, a perfect lattice with the lowest formation energy is dominant under the experimental conditions. The Kinetic effects can only play the role of smoothing the probability distribution of GBs, the extent of which depends on experimental conditions. Due to the kinetic effects, high probabilities of GBs near 15° (45) and 28° (32°) are reduced, and the rest of GBs with low probabilities are enhanced as shown in Fig. 4b. Especially, the probabilities of GB angles closing to 15° (16° ~ 19°) and 28° (23° ~ 25°) have obvious enhancement compared to (19° ~ 23°) which are far away from these two angles. This means that kinetic effects make some parts of GBs with high probabilities transform into their nearby GBs with low probabilities. This part of the study indicates that the kinetic effects tend to make the occurrence probability of GBs uniformly distribute in angle space, but the thermodynamic effects determined by grain-boundary energy still play the dominant role in the formation of GBs.

In summary, we have proposed an algorithm to characterize energetically favorable GBs by decomposing them into symmetric sub-GBs in two-dimensional materials. In this way, we have obtained an analytical energy functional for grain-boundary energy which only depends on the parameters of grain-boundary angle $\theta_M$ and asymmetric angle $\theta_L$. The decomposing algorithm is numerically verified by first principles calculations. Based on the energy functional, we have successfully mapped out the landscape of the grain-boundary energy in graphene, which is consistent with the high-throughput computation results. Furthermore, for the first time, we obtain the kinetic effects on the formation of GBs. We find that the formation of GBs in graphene is essentially controlled by the thermodynamics, and kinetic effects are somehow homogenizing the probabilities of GBs. The present work successfully presents an example for analytical energy functional of GBs, which would shed light on understanding the general properties of GBs and kinetic effects in other two-dimensional materials, such as BN, $MoS_2$, Mxene and even in metals.

**Notes**

The authors declare no competing financial interest.

**ACKNOWLEDGMENTS**

This work was supported by the Major Projects of National Natural Science Foundation of China (NSFC) (No. 12004100, 11991061) and National Key Research and Development Plan (No. 2016YFA0301001).